# Near-Field Thermal Energy Conversion by Tunneling to a Waveguide

Jacob L. Poole[*], Paul R. Ohodnicki

National Energy Technology Laboratory, 626 Cochrans Mill Road, Pittsburgh, PA 15236, USA

Energy is a vital resource and hence there is a continuous strive to improve upon existing technologies and to find new ones that address that basic need. The conversion of thermal energy is the primary method of generating electrical energy from a broad range of sources, for example fossil fuels, solar thermal, geothermal, and nuclear energy. A common need in all cases is the ability to efficiently extract the generated electromagnetic and thermal energy and to convert it to electricity. The current methods of thermal energy extraction are based on heat engines, thermoelectric and thermophotovoltaic conversion systems. In this report a method based on the direct extraction of Electromagnetic energy from the thermal near-field through tunneling and subsequent waveguiding, is presented.

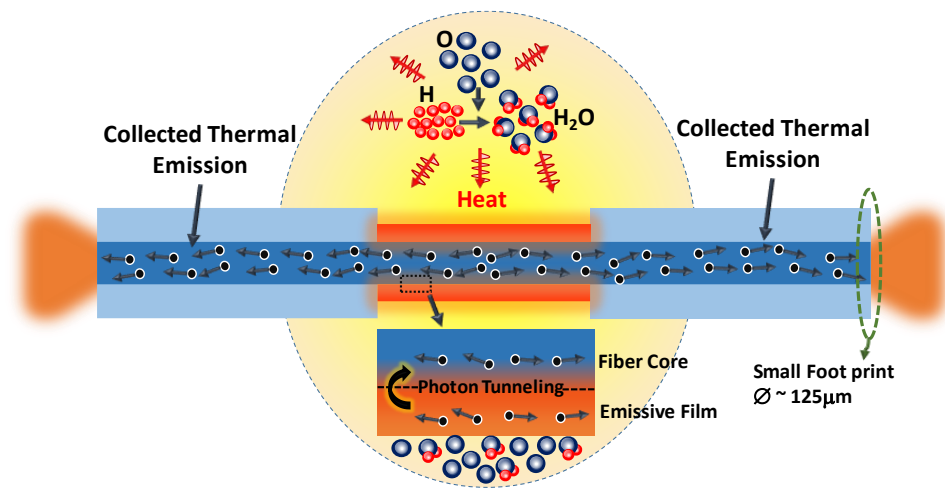

**Figure 1.** Schematic illustrating the extraction of thermal energy in the form of light by integrating functional thin-films with optical fiber.

Plank, after introducing his blackbody theory, himself has stated that it is only valid at large distances from surfaces, as he by then had already deduced that it was more complex below a certain distance. The theory of Fluctuation Electrodynamics by Rytov provided the tools to examine what happens to heat at distances very close to surfaces, thus predicting the thermal near-field, which exists at distances much less than the thermal emission wavelength ($\lambda_{Wien}$).[1-2] All matter above absolute zero possess thermal energy which is manifested in the vibrations of atoms, molecules, electrons, etc. This thermal energy introduces random displacements in the positions of charged particles generating random currents, which in turn produce electromagnetic fields.[3-8] Surfaces of materials can be charge separated acquiring surface plasmon polariton and surface phonon polariton resonant properties which have degrees of freedom, allowed k-vectors, parallel to the surface.[3, 6] As such, surface charge density waves cannot be observed at distances much greater that the thermal emission wavelength due to their evanescent nature, as they decay exponentially. Which may also explain the low thermal emissivity of materials in certain frequency bands, due to plasmon excitations.[6] It is not until some method to allow coupling of the surface k-vectors that the thermal emission near-field can be observed. Thus far, the primary method of observation has been through scattering techniques such as scanning near-field optical microscopy (SNOM), relying on bringing a tip near a surface under study where the tip scatters the near-field electromagnetic radiation into the far-field for detection.[4, 6, 9-10] On the other hand, large core optical fibers possessing thousands of optical modes are natural candidates for observing the thermal near-field, since the large mode density supplied by the fiber provides substantial k-vector components parallel with the near-field (**Figure 1**). Thus, optical fiber and other types of waveguides have the potential to provide high optical channel densities for extracting the thermal near-field electromagnetic radiation.[11]

It is predicted and experimentally verified that the thermal near-field has the potential to increase radiative heat transfer by orders of magnitude in comparison to the far-field.[4, 6, 8, 12] However, prior demonstrations were of high complexity through such techniques as SNOM and STM, which are not well suited for practical use. In contrast, the present topic of this report is the successful extraction of electromagnetic radiation from the thermal near-field with simple devices consisting of nothing more than thin-films integrated with optical fiber. To demonstrate the potential for extracting high power density electromagnetic radiation from the thermal near-field using an optical fiber waveguide, thin films of various oxides such as $TiO_2$, Nb-doped $TiO_2$, porous $TiO_2$, and a thermoelectric material CaMnNbO were examined. Preliminary results with these material systems demonstrate that as much as 50mWcm$^{-2}$ power density can be attained when the thin-film fiber composite was heated to 1100C. Furthermore, it is projected that at higher temperatures, such as the burning temperature of Hydrogen and Methane, power densities approaching as much as 5Wcm$^{-2}$ may be possible. These numbers are obtained with a heat source emitting radiation nearly orthogonal to the surface of the thin-films, as such, it would be expected that greater densities would be achieved when sources would emit radiation in parallel.

**Results and Discussions**

A series of measurements were performed to analyze the thermal energy extraction capability of optical fiber integrated with various thin films. **Table 1** summarizes the measurements where the films were first heated to 800°C in nitrogen containing 2% $O_2$. After measuring the optical power extracted by one end of the fiber, the chemistry of the environment was changed to 5% $H_2$ in $N_2$ and the extracted powers were reexamined. While holding the chemical composition, the furnace was heated to 1100C and the extracted optical power was reexamined. The table entry bare-fiber-end is a configuration in which a cleaved fiber end was placed against an $Al_2O_3$ block. Whereas, the configuration bare-fiber-through is when a piece of fiber without any processing was fed through to provide a reference for the measurements. The table entry labeled $CaMnNbO_3$ is when the fiber is coated 4 times with a solution yielding the thermoelectric material $CaMn_{0.98}Nb_{0.02}O_{3-\delta}$. The other thin-films were prepared by coating 4 times to yield film thicknesses on the order of 80nm. In all cases except for the thermoelectric material, as it is not stable in $H_2$, the exchange of $O_2$ with $H_2$ in the environment yielded a substantial increase in the extracted optical powers, as can be observed in **Table 1**.

**Table 1.** Measured optical power of the optical fibers coated with thin-films of various materials

|              | 800C (2%O2) | 800C(5%H2) | 1100C(5%H2) |
|--------------|-------------|------------|-------------|
| Nb-TiO2      | 4.25µW      | 5.15µW     | 24.15µW     |
| TiO2         | 3µW         | 4.3µW      | 22.6µW      |
| Mesoporous TiO2 | 2µW      | 3µW        | 21.75µW     |
| Bare-Fiber-End | 1.6µW     | 2.25µW     | 11.3µW      |
| Bare-Fiber-Through | 0.11µW | 0.3µW   | 1.37µW      |
| Etched-Fiber | 0.6µW       | 0.9µW      | 2µW         |
|              |             |            |             |
|              |             |            | 1100C(air)  |
| CaMnNbO3     |             |            | 20µW        |

In **Table 2** the extracted electromagnetic power density is calculated using the fiber geometry shown, where the diameter of the optical fiber is 125µm. Here, the extracted optical power was divided by the cross-sectional area of the fiber. Nb-TiO$_2$ yielded the best results providing a substantial power density of 50mWcm$^{-2}$. Whereas, the second-best measurement was obtained with TiO$_2$. Nb doping TiO$_2$ is known to provide a substantial increase in the free electron concentration while retaining moderate mobility values.[15]

**Table 2.** Calculated extracted power density at 1100C for the fiber dimensions shown.

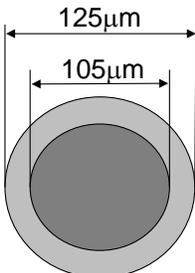

|              | Power Density (1100C) |
|--------------|-----------------------|
| Nb-TiO2      | 50mW/cm2              |
| TiO2         | 46mW/cm2              |
| Mesoporous TiO2 | 44mW/cm2           |
| CaMnNbO3     | 41mW/cm2              |
| Bare-Fiber-End | 23mW/cm2            |
| Etched Fiber | 3.36mW/cm2            |
| Bare-Fiber-Through | 2.8mW/cm2       |

In recent measurements it was confirmed that the electrical conductivity of an engineered TiO$_2$ film on fiber is approaching an astonishing value of 700Scm$^{-1}$ upon exposure to hydrogen at 800C. A conductivity value which support claims put forth in a previous report which were derived purely from optical measurements.[11] Therefore, in both cases a significant free carrier concentration exists while the mobility values remain in the moderate to low end. In TiO$_2$ hydrogen is expected to reduce the film generating oxygen vacancies in addition to interstitial Ti. Furthermore, hydrogen is also expected to interact with the generated defects in TiO$_2$ creating a substantial trap state population, which are thermally excitable.[16-18] Therefore, the following argument is proposed to explain the observed behavior. Because surface plasmon resonance was not observed for these films when probed with an NIR source having wavelengths up to 2200nm, the lifetimes of the electrons must be smaller than needed to create a sustained surface charge density wave. Otherwise, plasmon resonance would have been observed. However, the short lifetimes could mean that more photons are radiated away, which, in turn, means more extracted electromagnetic energy. Therefore, we term this as frustrated surface plasmon resonance which would have a substantial near-field emission and since the film thickness is on the order of 70nm, the near-field thermal emission can tunnel into the high optical mode density provided by the large core fiber.[11] On the other hand, mesoporous TiO$_2$ provided the third highest measured optical power, which is attributed to a substantial increase in the surface area in addition to TiO$_{1-x}$ defect interaction with hydrogen. Surface adsorbates can also increase the thermal emission by supplying excitable vibrational modes.[8] The fourth highest optical power density was provided by the thermoelectric material, which was a surprising result. However, thermoelectric materials are optimized for heat to electron flow conversion, not for heat to photon extraction (phonon to electron vs. phonon to photon). Pendry has shown that the maximum heat flux in a single channel is linked to the flow of information or to the flow of entropy, and from this conductivity values that maximize the heat transfer in a given channel and at a given temperature can be derived.[8] A simple prediction using this argument places the conductivity value to ~25Scm$^{-1}$ at 800C.[8] The reported conductivity value of Nb-TiO$_2$ is in this range at room temperature but it is expected to increase at higher temperatures.[15] Whereas the measured conductivity value of TiO$_2$ at higher temperatures is ~700Scm$^{-1}$. Therefore, the combination of high electron density and low mobility of Nb-TiO$_2$ is believed to be the reason why it performs better at extracting electromagnetic radiation from the thermal near-field.

**Table 3.** Measured and projected powers and power densities of mesoporous TiO$_2$ up to 2000C.

|  | Mesoporous TiO2 | | |
|---|---|---|---|
|  | Temperature | Power | Power Density |
| Measured | 800C (2%O2 in N2) | 2μW | 4.1mW/cm2 |
|  | 800C(H2) | 3μW | 6.1mW/cm2 |
|  | 900C(N2) | 5.5μW | 11.2mW/cm2 |
|  | 1000C(N2) | 10μW | 20.4mW/cm2 |
|  | 1100C(N2) | 17.5μW | 35.7mW/cm2 |
|  | 1100C(4%H2 in N2) | 21.75μW | 44.3mW/cm2 |
| Projected | 1200C(4%H2 in N2) | 39μW | 79mW/cm2 |
|  | 1300C(4%H2 in N2) | 70μW | 142mW/cm2 |
|  | 1400C(4%H2 in N2) | 126μW | 256mW/cm2 |
|  | 1500C(4%H2 in N2) | 227μW | 462mW/cm2 |
|  | 1700C(4%H2 in N2) | 409μW | 833mW/cm2 |
|  | 1800C(4%H2 in N2) | 735μW | 1497mW/cm2 |
|  | 1900C(4%H2 in N2) | 1323μW | 2695mW/cm2 |
|  | 2000C(4%H2 in N2) | 2381μW | 4850mW/cm2 |

With mesoporous TiO$_2$ a detailed series of measurements were obtained from 800°C to 1100°C, incrementally. The exchange of O$_2$ with H$_2$ yielded a 50% increase in the extracted optical power at 800°C. From the trends between 800°C and 1100°C a multiplier of 1.8 was obtained, meaning that the power at 100°C higher can be obtained from the previous value by a multiplication of 1.8. Using this trend, the extracted optical powers that would be measured at higher temperatures were estimated up to 2000°C, which is the temperature of combusting CH$_4$ and H$_2$. It is estimated that an astonishing power density of 4.8Wcm$^{-1}$ may be possible when extracted with optical fiber type waveguides. Although, to achieve those values the material constituents would need to be those that can survive such high temperatures. The 1100C that the current measurements were taken at is actually the upper limit of silica fibers and the material can start to deform. However, there are numerous potential material candidates such as sapphire that can withstand such extreme temperatures.

An Nb-TiO₂ film of thickness ~40nm was further characterized by examining the optical spectrum of the extracted optical power (**Figure 2**). After heating the processed fiber fastened inside the quartz tube furnace to 800°C in air the solid line spectra labeled with 1.9µW was obtained which, after exchanging the O₂ with H₂ yielded a power increase of ~124% from 1.9µW to 4.25µW, whereas only a ~44% increase was noted at 1100°C.

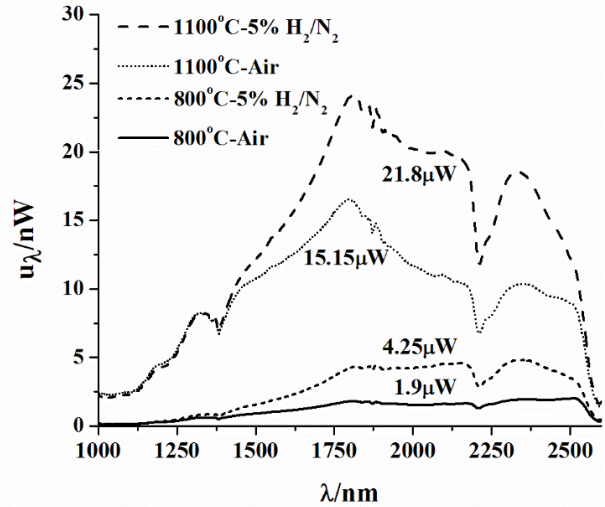

**Figure 2.** Optical spectrum of the optical fiber extracted electromagnetic energy with an Nb-TiO₂ thin-film of ~40nm thickness at 800C and 1100C with O₂ and when the O₂ was exchanged with H₂.

In addition, the thermally generated optical power of specialty erbium doped fiber (ER100-70/235DC, NA ~0.115, NLIGHT) was measured. For characterization, a 10cm piece of fiber was fusion spliced to a 400um core 0.22 NA fiber for collection and guiding to the power meter. The optical power was measured in nitrogen and at 800°C and a cross-sectional power density of 83mWcm$^{-2}$ was obtained, which increased to 408mWcm$^{-2}$ at 1100°C. It is important to note that the NA of this fiber is ~0.115, and that the power collected by a fiber is proportional to NA to the 4$^{th}$ power. Therefore, increasing the NA should have a drastic effect on the thermal light collection capability of fiber type waveguides. An Erbium doped fiber emitter is better suited for this configuration, in which thermal energy is emitted mostly orthogonal to the fiber. This, in comparison with a plasmonic film which would be better excited by thermal energy with a momentum vector that is parallel with the fiber's surface. Therefore, more elaborate schemes would need to be constructed in order to compare the thermal energy harvesting capability of the different types of fibers presented here. But, the erbium doped fiber is much better suited to capture orthogonally radiated heat, as in this case, as opposed to a film having plasmonic behavior.

**Table 4.** Measured optical power and calculated cross-sectional power density of a 10cm section of erbium-doped fiber (ER100-70/235DC, NLIGHT).

| Erbium Doped Fiber | | |
|---|---|---|
| Temperature | Power | Power Density |
| **800C** | 36µW | 83mW/cm2 |
| **900C** | 71µW | 164mW/cm2 |
| **1000C** | 106µW | 244mW/cm2 |
| **1100C** | 177µW | 408mW/cm2 |

It would be interesting to explore materials systems in which the thermal emission wavelength coincides with the metal insulator transition, and at the Currie point, or even a mixture of both. Other interesting materials would be clathrates which have high electronic conductivity but block the transfer of heat, and Heusler type materials which accomplish the same. Where, the general idea is the maximization of the phonon to photon conversion efficiency at the order do disorder cyclical transitions. This could yield thermal energy powered pulsed light sources, as one example.

**Conclusions**

In this report a waveguide-based method to extract electromagnetic radiation from the thermal near-field is presented, which may be for power generation for applications in cooling. Once the thermal energy is captured in the form of light in the waveguide, it is propagated with little loss to arbitrary distances. By integrating thin-films of highly functional materials such as $TiO_2$ and $Nb-TiO_2$ with optical fiber, the extracted electromagnetic power density reached 50mWcm$^{-2}$, which is substantial and compares well with existing methods of thermal energy conversion. Projecting the observed trends in the extracted near-field electromagnetic radiation from 1100C to 2000C, which is near the combusting temperature of $CH_4$ and $H_2$, it is estimated that a power density approaching 5Wcm$^{-2}$ could be potentially achieved. However, the material constituents would need to be replaced with ones capable of operating at such high temperatures. The erbium doped fiber tested yielded much stronger power densities, but that fiber is much better suited to converting orthogonally radiated thermal energy into light in the fiber, than a thin-film using by plasmon resonance. However, erbium-doped fiber is much simpler to manufacture and it highly useful. It is hoped that these results will spur further exploration of this energy conversion topic!

**Acknowledgement**